# Investigation of stimulated dynamics in strongly anisotropic high-temperature superconductors system Bi-Pb-Sr-Ca-Cu-O.

J.G. Chigvinadze*, S.M. Ashimov*, T. V. Machaidze*

*  *Andronikashvili Institute of Physics, 0177, Tbilisi, Georgia*

**Abstract**

It is used the mechanical method of Abrikosov vortex stimulated dynamics investigation in superconductors. With its help it was studied relaxation phenomena in vortex matter of high-temperature superconductors. It established that pulsed magnetic fields change the course of relaxation processes taking place in vortex matter. The study of the influence of magnetic pulses differing by their durations and amplitudes on vortex system of strongly anisotropic high-temperature superconductors system $Bi_{1.7}Pb_{0.3}Sr_2Ca_2Cu_3O_{10-\delta}$ showed the presence of threshold phenomena. The small duration pulses doesn't change the course of relaxation processes taking place in vortex matter. When the duration of pulses exceeds some critical value (threshold), then their influence change the course of relaxation process which is revealed by stepwise change of relaxing mechanical moment $\tau^{rel}$.

These investigation showed that the time for formatting of Abrikosov vortex lattice in $Bi_{1.7}Pb_{0.3}Sr_2Ca_2Cu_3O_{10-\delta}$ is of the order of 150 μs which on the order of value exceeds the time necessary for formation of a single vortex observed in isotropic high-temperature superconductor $HoBa_2Cu_3O_{7-\delta}$ and on two orders exceeds the creation time of a single vortex observed in classical type II superconductors.

## 1. Introduction

The present work is devoted to the experimental investigation of Abrikosov vortex lattice dynamics in strongly anisotropic high-temperature superconductors of Bi-Pb-Sr-Ca-Cu-O system using stimulating pulsed magnetic fields.

Investigation of vortex matter stimulated dynamics in strongly anisotropic high-temperature superconductors is one of important problems both for understanding of fundamental problems connected with strongly anisotropic high-temperature superconductors [1], and from the point of view of their practical applications particularly their behavior in electromagnetic fields and under the current loading [2].

The critical temperature of this class of high-temperature superconductors is so high that they remain superconductive at temperatures when thermal fluctuations play essential role because their energy becomes compared with the elastic energy of vortex and, as well, with the pinning energy [3]. It creates prerequisites for phase transitions. Due to the layered crystal structure and anisotropy, which are characteristic of high-temperature superconductors, it is created conditions for the appearance on *B-T* diagram.( B is magnetic induction, T-is temperature) different phases [4-15].

One of fine examples of a phase transition in vortex matter is the *3D* tree-dimensional transition of Abrikosov vortices in *2D* quasi-two-dimensional ones, so-called Pancake vortices. Such phase transition takes place in high-temperature superconductor system $Bi_{1.7}Pb_{0.3}Sr_2Ca_2Cu_3O_{10-\delta}$ [12,13]. During this transition a sharp dissipation energy decrease of moving Abrikosov vortices (almost two orders of value) takes place what in its turn could be related with an essential decrease of pinning force. Further experiments showed out that in the same material the pinning force also sharply increases at the *3D-2D* transition (approximately on 300% in value) [15] what makes such materials perspective for technical applications, all the more so that the upper critical field $H_{c2}$ when superconductivity is destroyed could reach 150 T [1]. It is essentially higher as compared with $H_{c2}$ in traditional type II superconductors used currently in practice.

Investigations of Abrikosov vortex lattice dynamics stimulated by alternative and pulsed magnetic fields in high-temperature superconductors could undoubtedly play decisive role from the point of view at their fitness for practical applications in devices under construction operating on the basis of high-temperature superconductivity because frequently alternating and pulsed magnetic fields put limits on the operation of aforementioned devices and such fields, as a rule, appear at their operation. Therefore the investigation of influence of pulsed and alternating magnetic fields on Abrikosov vortex lattice dynamics and relaxation processes occurring in vortex matter is very important.

At investigation of Abrikosov vortex lattice stimulated dynamics on the external permanent magnetic fields, applied to HTSC and creating vortex lattice, is imposed a weak alternative or pulsed magnetic field, causing the motion of vortex continuum. This, in its turn, leads to a change of relaxation processes taking place in the vortex matter [8, 9].

Consequently, the study of these problems could results in the understanding of energy dissipation mechanisms arising at the motion of vortex matter inside a superconductor. Investigations of these problems is of current concern from the practical view because it makes it possible to establish of applicability limits of technical devices constructed and created on the basis of high-temperature superconductors.

## 2. EXPERIMENTAL

For investigation it was used currentless mechanical method of Abrikosov vortex stimulated dynamics study by magnetic pulses revealing relaxation phenomena in vortex matter described in work [16]. This method is a development of currentless mechanical method of pinning investigations[17,18] and is based on pinning forces countermoments measurements and viscous friction, acting on a axially symmetrical superconducting sample in an outer (transverse) magnetic field. Countermoments of pinning forces and of viscous friction, acting on a superconductive sample from quantized vortex lines side (Abrikosov vortices) are defined the way as it was described in work [19,20]. The sensitivity of the method accordingly work[21], is equivalent to $10^{-8}$ V×cm$^{-1}$ in the method of *V-A* characteristics.

The high-temperature superconducting samples of $Bi_{1.7}Pb_{0.3}Sr_2Ca_2Cu_3O_{10-\delta}$ system were prepared by the standard solid state reaction method. Samples were made cylindrical with height L=12mm and diameter d=1.2mm. Their critical temperature was Tc=107 K. The investigated samples were isotropic what was established by mechanical moment $\tau$ measurements appearing $H > H_{c1}$ with the penetration of Abrikosov vortices into a freely suspended on a thin elastic thread superconducting sample. The appearance of such moment $\tau = MH\sin\alpha$, characteristic for anisotropic superconductors, is related with penetrating Abrikosov vortices and the mean magnetic moment $\vec{M}$ of a sample which could deviate on angle $\alpha$ from the direction of outer magnetic field $\vec{H}$. In superconducting anisotropic samples it is presented energetically favorable directions for the arrangement of emerging (penetrating) vortex lines which in their turn are fastened by pinning centers creating aforementioned moment $\tau$. The lack of $\tau$ moment is characteristic for isotropic and investigated by us samples, no matter magnetic field value and its previous orientation in respect to $\vec{H}$ in the axial symmetry plane. Pulsed magnetic fields were created by Helmholtz coils. The value of pulsed magnetic fields was changed in $\Delta h = 2 \div 200 Oe$ limits.

In experiments it was used both single and continuous pulsed with repetition frequency ν from 2.5 s$^{-1}$ to 500s$^{-1}$. The duration Δx of pulses was changed from 0,5 до 500 μs. Magnetic pulse could be directed both parallel *(Δh||H)* and perpendicularly *(Δh⊥H)* to applied steady magnetic field $\vec{H}$, creating mixed state of superconducting sample. The standard pulsed generator and amplifier were used to feed Helmholtz coils. The current strength in coils reached up to 40÷50 A. Samples were high-temperature superconductors of $HoBa_2Cu_3O_{7-\delta}$ system placed in the center between Helmholtz coils.

The principal set-up of experiment is shown in fig.1 [19,20]. In experiments it is measured the rotation angle $\varphi_2$ of sample depending on the angle of rotation of a torsion head $\varphi_1$, transmitting the rotation to a sample by means of suspension having the torsion stiffness $K \approx 4\cdot10^{-8}$ N m, which can be replaced when necessary by a less stiff or stiffer one.

The measurements were carried out at a constant speed of rotation of the torsion head, making ω$_1$=1,8·10$^{-2}$ rad/s . Angles of rotation $\varphi_1$ and $\varphi_2$ were determined with an accuracy of ±4,6·10$^{-3}$ and ±2,3·10$^{-3}$ rad, respectively. The uniformity of the magnetic field's strength along a sample was below $\Delta H / H = 10^{-3}$.

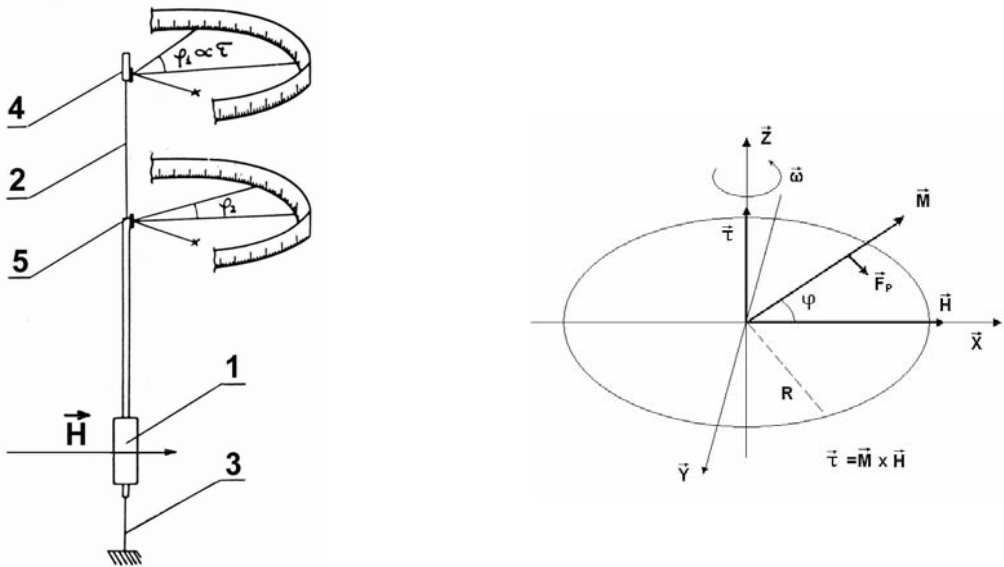

Fig. 1. The schematic diagram and the geometry of the experiment. 1 - sample, 2 - upper elastic filament, 3 - lower filament, 4 - leading head, 5 - glass road. φ is angle between $\vec{M}$ and $\vec{H}$.

To avoid effects, connected with the frozen magnetic fluxes, the lower part of the cryostat with the sample was put into a special cylindrical Permalloy screen, reducing the Earth magnetic field by the factor of 1200. After a sample was cooled by liquid nitrogen to the superconducting state, the screen was removed, a magnetic field of necessary intensity $H$ was applied and the $\varphi_2(\varphi_1)$ dependences were measured. To carry out measurements at different values of $H$, the sample was brought to the normal state by heating it to до $T > T_c$ at $H = 0$, and only after returning sample and torsion head to the initial state $\varphi_1 = \varphi_2 = 0$, the experiment was repeated.

## 3. Results and discussions.

During rotation of the sample both of normal and superconducting states in the absence external magnetic field ($H = 0$) the $\varphi_2$ dependence versus $\varphi_1$ is linear and the condition is satisfied.

$$\varphi_1 = \varphi_2 = \omega t$$

The character of the $\varphi_2(\varphi_1)$ dependence is changed significantly, when the sample is in magnetic fields $H > H_{c1}$ at $T < T_c$. Typical $\varphi_2(\varphi_1)$ dependences at T=77K and various magnetic fields for HoBa$_2$Cu$_3$O$_{7-\delta}$ sample ( length of a cylindrical sample L=13mm and diameter d=6mm ) is shown in Fig.2. [22], (such dependence in strongly anisotropic superconductors is only observed in relatively weak fields close to $H \geq H_{c1}$ ).

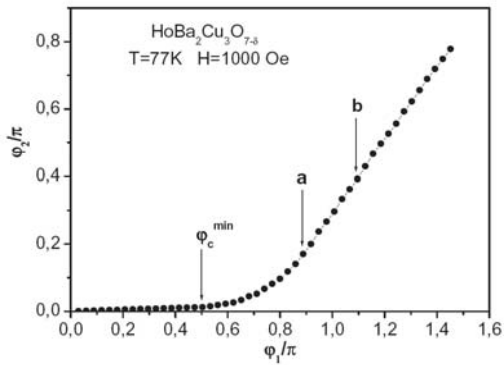

Fig.2. Dependence of the rotation angle of the sample HoBa$_2$Cu$_3$O$_{7-\delta}$ $\varphi_2$ on the rotation angle of the leading head $\varphi_1$ in magnetic field H=1000 Oe at T=77K.

Three distinct regions are observed in Fig.2. In the first (initial) region, the sample does not respond to the increase in $\varphi_1$, i.e. to the applied and increase with time torsion torque as $\varphi_1 \sim \tau = K(\varphi_1 - \varphi_2)$ or responds weakly. Such behaviour of the sample can be explained by fact that Abrikosov vortices are not detached from pinning centers at small values of $\varphi_1 \sim \tau$, but if the sample is still turned slightly, this can be caused by elastic deformation of magnetic force lines beyond it or, possibly, by separation of the most weakly fixed vortices. As it is seen from Fig.2, as soon as a certain critical value $\varphi_c^{min}$ depending on $H$ is reached, the first region under goes a transition to the second region in which the velocity of the sample increases gradually with $\varphi_1$ increasing resulting from the progressive process of detachment of vortices from their corresponding pinning centers. One should expect that just in this region, in the rotating sample "the vortices fan" begins to unfold, in with the vortices are distributed according to the instantaneous angles of orientations with respect to the fixed external magnetic field. In this case the of orientation angles of separate vortex filaments are limited from $\varphi_{fr}$ to $\varphi_{fr} + \varphi_{pin}$, where $\varphi_{fr}$ is the angle on which the vortex filament can be turned with respect to $\vec{H}$ by forces of viscous friction with the matrix of superconductor, and $\varphi_{pin}$ is the angle on with the vortex filament can be turned by the most strong pinning center, studied for the first time in [23].

The gradual transition (at high $\varphi_1$ values) to the third region where the linear $\varphi_2(\varphi_1)$ dependence was observed, allows one to define the countermoments of pinning forces $\tau_p$ and $\tau_{fr}$, independently. Just in this region, when $\omega_1 = \omega_2$ the torque $\tau$, appeared to the uniformly rotating sample, is balanced by the countermoment $\tau_p$ and $\tau_{fr}$. In particular, in the case of continuously rotating sample with frequency $\omega_1 = \omega_2$ one could find similarly to [24, 25] the expression for the total braking torque $\tau$ [19].

Indeed, if we consider in this case a vortex element $d\vec{s}$ moving with velocity $\vec{\upsilon}_\perp$ perpendicular to $d\vec{s}$, then the average force acting on this elements is

$$d\vec{f}_\upsilon = \vec{\upsilon}_\perp \eta ds + \frac{\vec{\upsilon}_\perp}{|\upsilon_\perp|} F_l ds$$

and the associated braking torque, exerted on the rotating specimen becomes:

$$d\vec{\tau} = \vec{r} \times d\vec{f}_\upsilon$$

where $\vec{r}$ is the vector pointing from the rotational axis to the vortex elements, $F_l$ is the pinning force per flux thread per unit length, and $\eta$ is the viscosity coefficient. For a cylindrical specimen of radius $R$ and height $L$ integrating over the individual contribution of all vortex gives a total braking torque $\tau$

$$\tau = \tau_p + \tau_0 \omega \qquad (1)$$

with

$$\tau_p = \frac{4}{3}\frac{BF_l}{\Phi_0}LR^3 \ ,$$

and

$$\tau_0 = \frac{\pi}{4}\frac{B}{\Phi_0}\eta LR^4 \ ,$$

Where $B$ is the inductivity averaged over the sample, $\Phi_0$ is the flux quantum, $L$ is the height and $R$ is the radius of the sample.

As it is shown in Fig.2, starting with the point (**a**), where $\omega_1 = \omega_2$, to the superconducting sample uniformly rotating in the homogeneous stationary magnetic field H=1000 Oe, is applied stationary dynamic torsion moment $\tau_p^{dyn} = \tau_p^{st} + \tau_{fr}$.

If in this region the torsion head is stopped, then at the expense of relaxation processes connected with the presence of viscous forces acting on vortex filaments, the sample will continue the rotation in the same direction (with decreasing velocity) until it reaches a certain equilibrium position, depending on the $H$ value. The Fig.3 shows curves of $\tau^{rel}$ time dependences at the stopped leading head for $Bi_{1.7}Pb_{0.3}Sr_2Ca_2Cu_3O_{10-\delta}$ sample at T=77K and H=500 Oe.

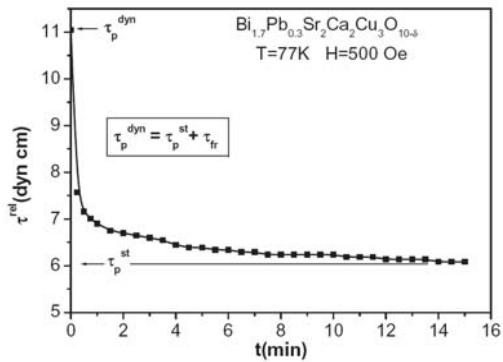

Fig.3. Dependence of momentum $\tau^{rel}$ on time $t$ after the stopping of rotating head for $Bi_{1.7}Pb_{0.3}Sr_2Ca_2Cu_3O_{10-\delta}$ sample at T=77K and H=500 Oe.

If during the relaxation after rotation of sample one applies the pulsed magnetic field in parallel to the outer magnetic field $\vec{H}$, then additional vortices, created as result of magnetic pulse, influence the structure already existing in the sample as "the vortex fan" what could result in the decrease of the angle of its unfolding or to its folding. The letter in its turn, would cause the additional change in the relaxation process taking place in the sample, and, correspondingly, results in the stepwise decrease of moment related with viscous forces $\tau_{fr}$.

But the change of relaxation process character and, correspondingly, the stepwise decrease of moment could happen if the duration of magnetic pulse is larger as compared with the time necessary for creation of a new vortex structure, which will influence the superconducting sample relaxing in magnetic field. If it is the case, then at the small durations of magnetic pulses the relaxation curve, presented in Fig.3, doesn't change, but when this duration becomes the order of a time for penetration of vortices into the sample and the creation of vortex structure, then the aforementioned change of relaxation processes could principally appear.

In the presented work it was studied the influence of different duration and amplitude pulses on relaxation processes in vortex matter.

In Fig. 4 it is presented results of investigations of different duration magnetic pulses influence on relaxation processes in vortex matter. The application of magnetic pulses in all cases was made after the stopping of rotation head of the suspension system on the fifth minute of relaxation process.

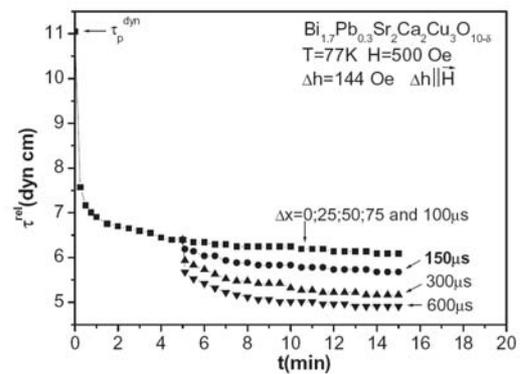

Fig.4. Dependence of momentum $\tau^{rel}$ on time $t$ with the influence at t=5 min on the relaxation process of single magnetic pulse $\Delta h=144$ Oe with durations $\Delta x=0 \div 600\mu s$. The pulsed magnetic field is parallel to the main magnetic field H=500 Oe applied to $Bi_{1.7}Pb_{0.3}Sr_2Ca_2Cu_3O_{10-\delta}$ sample at T=77K.

These results show at durations of magnetic pulses $\Delta x$=25, 50, 75 and 100 μs relaxation curves doesn't change what apparently is related with the fact that during the time duration of pulses Abrikosov vortices doesn't have enough time to enter into the bulk of superconductor. With the increase of duration of pulses starting from 150 μs and higher, the relaxation curve changes step wisely and relaxes further on some lower level, what is shown in Fig. 4. As it is seen from the figure the value of stepwise change of relaxing momentum increases with the increase of pulse duration.

At durations of magnetic pulses up 100μs their influence doesn't change the relaxation process, and at durations of appeared pulses >100μs it is observed the stepwise change of $\tau^{rel}$, what manifests on the exiting of threshold $\Delta x_c$. So, from results presented in Fig.4. one could conclude that the Abrikosov vortex system creation time in high-temperature anisotropic superconductor $Bi_{1.7}Pb_{0.3}Sr_2Ca_2Cu_3O_{10-\delta}$ makes the value of the order of 150 μs. This value almost on the order of value higher then the Abrikosov vortex lattice creation time in the isotropic high-temperature superconductor of $HoBa_2Cu_3O_{7-\delta}$ system which is of the order of 20μs[22] and on two orders of value higher then the time for creation of single vortex for the first time measured by G. Boato, G.Gallinaro and C. Rizzuto [26], who showed that this time is less then $10^{-5}$ sec. in classic type II superconductor.

The results presented in Fig.5 on the action of different durations single pulses on relaxation processes of vortex matter and, correspondingly, on the mechanical moment $\tau^{rel}$ clearly show the existence of the threshold $\Delta x_c$. As one could expect, the value of threshold duration $\Delta x_c$ doesn't depends on the strength of field applied in parallel to the main magnetic field.

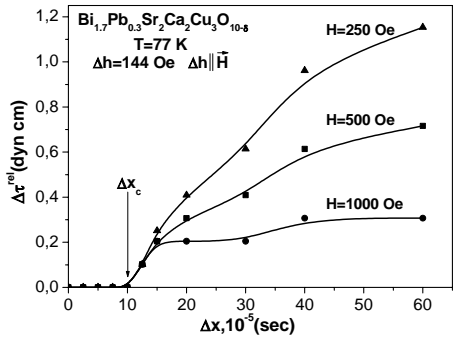

Fig.5. Dependence of momentum $\Delta \tau^{rel}$ on the duration $\Delta x$ of a single pulse of magnetic field $\Delta h=144$ Oe applied in parallel to the main magnetic field (H=250 Oe, H=500 Oe and H=1000 Oe) $Bi_{1.7}Pb_{0.3}Sr_2Ca_2Cu_3O_{10-\delta}$ sample at T=77K.

And, finally, in Fig. 6 and in Fig.7 it is presented the results of magnetic pulses influence on relaxation processes of strongly anisotropic high-temperature superconductor $Bi_{1.7}Pb_{0.3}Sr_2Ca_2Cu_3O_{10-\delta}$ vortex matter on different stages of the relaxation process. These results presented in Fig.6, also show that the critical duration of pulses is $\Delta x_c = 100\mu s$. For the creation of vortex structure it is necessary a time of order of 150μs.

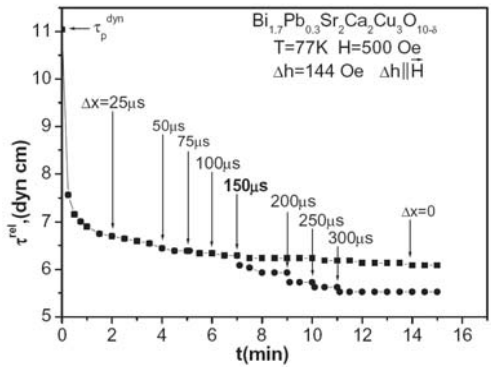

Fig.6. Dependence of momentum $\tau^{rel}$ on time $t$ with consecutive influence over some definite intervals on the relaxation process of single magnetic pulse $\Delta h=144$ Oe with increasing durations $\Delta x$. The pulsed magnetic field is parallel to the main H=500 Oe field applied to the sample $Bi_{1.7}Pb_{0.3}Sr_2Ca_2Cu_3O_{10-\delta}$ at T=77K.

But one should note that multiple application of increasing length single pulses more and more decrease the value of observed stepwise changes $\Delta \tau^{rel}$. This manifests on the decrease of the number of Abrikosov vortex threads relaxing in sample and their reattachment on more strong pinning centers with the decrees of "vortex fan's" unfolding angle.

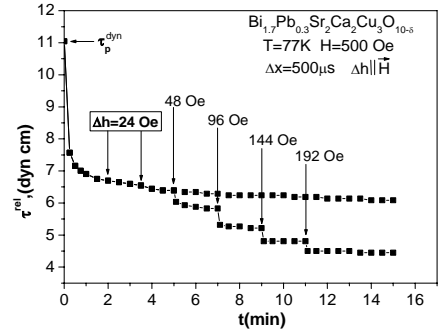

Fig.6. Dependence of momentum $\tau^{rel}$ on time $t$ at different values of pulsed magnetic field $\Delta h$, applied to the $Bi_{1.7}Pb_{0.3}Sr_2Ca_2Cu_3O_{10-\delta}$ sample at T=77K in a magnetic field H=500 Oe.

Results, presented in Fig. 7, show that the critical value of applied pulsed magnetic field is $\Delta h=24$ Oe. For changing of relaxation if is necessary a pulsed magnetic field of the order of 30 Oe what practically coincides with the first critical field $H_{c1}$, for $Bi_{1.7}Pb_{0.3}Sr_2Ca_2Cu_3O_{10-\delta}$ sample. But it should be stressed that we so far doesn't know whether this coincidence is accidental of it holds for all systems of high-temperature superconductors.

## 4. Conclusion

The simple mechanical method of Abrikosov vortex lattice stimulated dynamics investigation was applied to study the influence of pulsed magnetic fields on relaxation phenomena in vortex matter of strongly anisotropic high- temperature superconductors. It was observed the change of relaxation processes in vortex matter as result of influence on it by pulsed magnetic fields.

The study of different duration and amplitude pulsed magnetic fields influence showed the existence of threshold phenomena. The short duration magnetic pulses don't change the course of relaxation processes in vortex matter of strongly anisotropic high- temperature superconductor $Bi_{1.7}Pb_{0.3}Sr_2Ca_2Cu_3O_{10-\delta}$. When the duration of pulses is longer than some critical value (threshold), then their influence change the course of relaxation phenomena. The last is revealed by the stepwise change of relaxing mechanical moment $\tau^{rel}$, apparently, related with a sharp change of pinning and rearrangement of vortex system of superconducting sample when new vortex bunches enter into the its bulk at application of pulsed field in addition to the outer magnetic field creating the main vortex structure in the investigated. The new bunch "shakes" the existing vortex lattice in the sample causing detachment of vortices from a weak pinning centers and their reattachment on new centers what, apparently, is the reason of the stepwise decrease of mechanical momentum $\tau^{rel}$.

These made it possible to define the time of Abrikosov vortex lattice creation in strongly anisotropic high-temperature superconductor $Bi_{1.7}Pb_{0.3}Sr_2Ca_2Cu_3O_{10-\delta}$, which appeared to be on the order of value higher as compared with the time of creation in isotropic high-temperature superconductors $HoBa_2Cu_3O_{7-\delta}$ and on the two orders higher then the creation time of single vortex observed in classical type II superconductors.

**Acknowledgement**

The work was supported by the grants of International Science and Technology Center (ISTC) G-389 and G-593